\documentclass[prd,twocolumn,aps,showpacs]{revtex4}

\def\[{\left\lbrack}
\def\]{\right\rbrack}

\def\({\left(}
\def\){\right)}
\newcommand{\be}{\begin{equation}}
\newcommand{\ee}{\end{equation}}
\newcommand{\ea}{\end{eqnarray}}
\newcommand{\ba}{\begin{eqnarray}}

\newcommand{\vx}{{\vec{x}}}
\newcommand{\vy}{{\vec{y}}}

\newcommand{\ep}{{\epsilon}}

\newcommand{\dirac}{{\delta(\vx - \vy)}}

\def\f{\phi}
\def\p{\psi}

\def\ep{\epsilon}

\def\pa{\partial}
\def\l{\label}

\begin{document}

\title{The dual embedding method in $D=3$}

\author{E. M. C. Abreu$^a$\footnote{\sf E-mail: evertonabreu@ufrrj.br},A.C.R. Mendes$^b$\footnote{\sf E-mail: albert@fisica.ufjf.br},C. Neves$^b$\footnote{\sf E-mail: cneves@fisica.ufjf.br},W. Oliveira$^b$\footnote{\sf E-mail: wilson@fisica.ufjf.br}, F.I. Takakura$^b$\footnote{\sf E-mail: takakura@fisica.ufjf.br} and L. M. V. Xavier$^b$} 
\affiliation{${}^{a}$Departamento de F\'{\i}sica, Universidade Federal Rural do Rio de Janeiro\\
BR 465-07, 23851-180, Serop\'edica, Rio de Janeiro, Brazil\\
${}^{b}$Departamento de F\'{\i}sica,
ICE, Universidade Federal de Juiz de Fora,\\
36036-330, Juiz de Fora, MG, Brazil\\
\bigskip
\today}


\begin{abstract}
\noindent  Improving the beginning steps of a previous work, we settle the dual embedding method (DEM) as an alternative and efficient method for obtaining dual equivalent actions also in $D=3$. 
We show that we can obtain dual equivalent actions which were previously obtained in the literature using the gauging iterative Noether dualization method (NDM). We believe that, with the arbitrariness property of the zero mode, the DEM is more profound since it can reveal a whole family of dual equivalent actions.  After a review of our previous work, we obtain the dual equivalent theory of the self-dual model minimally coupled to $U(1)$ charged bosonic matter.  The result confirms the one obtained previously which is important since it has the same structure that appears in the Abelian Higgs model with an anomalous magnetic interaction.  
\end{abstract}
\pacs{11.15.-q,11.10.Ef,11.30.Cp}

\maketitle

\section{Introduction}

In current times we are living in an intense production of papers about issues concerning duality, which, in a nutshell, can be described as two equivalent versions of a model using different fields.  Actually, we can name several different contexts in theoretical physics in which duality is an essential ingredient \cite{hl}.

Using the well known equivalence between self-dual \cite{tpn} and the topologically massive models \cite{djt} proved by Deser and Jackiw 
\cite{dj} through the master action approach \cite{hl}, a correspondence has been established between the partition functions for the massive Thirring model and the Maxwell-Chern-Simons (MCS) theories \cite{fs}.  The situation for the case of fermions carrying non-Abelian charges, however, is less understood due to a lack of equivalence between these vectorial models, which has only been established for the weak coupling regime \cite{bfms}.  As critically observed in \cite{klrn} and \cite{bbg}, the use of master actions in this situation is ineffective for establishing dual equivalences.
The so-called gauging iterative Noether dualization method \cite{ainrw} has 
been shown to thrive in establishing some dualities between models \cite{iw}.  However, this
method provides a strong suggestion of duality since it has been shown to give
the expected result in the paradigmatic duality between the so-called self-dual model \cite{tpn} and the Maxwell-Chern-Simons 
theory in three dimensions duality.  This correspondence was first established by Deser and Jackiw \cite{dj} 
and using a parent action approach \cite{suecos}.

The symplectic embedding method  \cite{aon} is not affected by ambiguity problems.  It has the great advantage of being a simple and direct way of choosing the infinitesimal gauge generators of the built gauge theory.  This give us a freedom to choose the content of the embedded symmetry according to our necessities.  This feature makes possible a greater control over the final Lagrangian.  This method can avoid the introduction of infinite terms in the Hamiltonian of embedded non-commutative and non-Abelian theories.  This can be accomplished because the infinitesimal gauge generators are not deduced from previous unclear choices.  Another related advantage is the possibility of doing a kind of general embedding, that is, instead of choosing the gauge generators at the beginning, one can leave some unfixed parameters with the aim of fixing them later, when the final Lagrangian has being achieved.  Although one can reach faster the final theory fixing such parameters as soon as possible, this path is more interesting in order to study the considered theory, and it is helpful if the desired symmetry is unknown, but some aspects of the Lagrangian are wanted.  
This approach to embedding is not dependent on any undetermined constraint structure and also works for unconstrained systems.  This is different from all the existent embedding techniques that we use to convert \cite{batalin,wotzasek}, project \cite{vyth} or reorder \cite{mitra} the existent second-class constraints into a first-class system.  This technique on the other hand only deals with the symplectic structure of the theory so that the embedding structure does not rely on any pre-existent constrained structure.

In \cite{aon} two of us demonstrated that the DEM does not change the physical contents originally present in the theory computing the energy spectrum.  This technique follows Faddeev's suggestion \cite{FS} and is set up on a contemporary framework to handle noninvariant models, namely, the symplectic formalism \cite{FJ}.

In order for this work to be self-sustained, it is organized as
follows: In section II, we present a brief review of the dual embedding formalism. 
In section III, the $D=4$ massive Carrol-Field-Jackiw (CFJ) theory will be analyzed  from the
symplectic point of view \cite{FJ}. Here, the Dirac brackets
among the fields will be computed. In section IV, the DEM will be used in $D=3$ and, as a consequence, the
gauge-invariant/dual equivalent version of the self-dual model minimally coupled to $U(1)$ charged bosonic matter will be obtained. Finally, the conclusions and perspectives are accomplished in the last section.

\section{The dual embedding formalism}

As said in the last section, this technique follows the Faddeev-Shatashivilli's
suggestion \cite{FS} and is set up on a contemporary framework to handle constrained
models, i.e., the symplectic formalism \cite{FJ}.  
In the following lines, we try to keep this paper self-sustained reviewing the main steps of the dual embedding formalism.  We will follow closely the ideas contained in 
\cite{amnot}.

Let us consider a general noninvariant mechanical model whose dynamics is governed by a Lagrangian
${\cal L}(a_i,\dot a_i,t)$, (with $i=1,2,\dots,N$), where $a_i$ and $\dot a_i$
are the space and velocities
variables, respectively. Notice that this model does not result in the loss of
generality nor physical content. Following the symplectic method the zeroth-iterative
first-order Lagrangian one-form is written as
 \begin{equation}
\label{2000}
{\cal L}^{(0)}dt = A^{(0)}_\theta d\xi^{(0)\theta} - V^{(0)}(\xi)dt,
\end{equation}
and the symplectic variables are
\be
\xi^{(0)\theta} =  \left\{ \begin{array}{ll}
                               a_i, & \mbox{with $\theta=1,2,\dots,N $} \\
                               p_i, & \mbox{with $\theta=N + 1,N + 2,\dots,2N ,$}
                           \end{array}
                     \right.
\ee
where $A^{(0)}_\theta$ are the canonical momenta and $V^{(0)}$ is the symplectic potential. From the Euler-Lagrange equations of motion, the symplectic tensor is obtained as
\begin{eqnarray}
\label{2010}
f^{(0)}_{\theta\beta} = {\partial A^{(0)}_\beta\over \partial \xi^{(0)\theta}}
-{\partial A^{(0)}_\theta\over \partial \xi^{(0)\beta}}.
\end{eqnarray}
If the two-form $f \equiv \frac{1}{2}f_{\theta\beta}d\xi^\theta \wedge d\xi^\beta$ is singular, the symplectic matrix (\ref{2010}) has a zero-mode $(\nu^{(0)})$ that generates a new constraint when contracted with the gradient of the symplectic potential,
\begin{equation}
\label{2020}
\Omega^{(0)} = \nu^{(0)\theta}\frac{\partial V^{(0)}}{\partial\xi^{(0)\theta}}.
\end{equation}
This constraint is introduced into the zeroth-iterative Lagrangian one-form Eq.(\ref{2000}) through a Lagrange multiplier $\eta$, generating the next one
\begin{eqnarray}
\label{2030}
{\cal L}^{(1)}dt &=& A^{(0)}_\theta d\xi^{(0)\theta} + d\eta\Omega^{(0)}- V^{(0)}(\xi)dt,\nonumber\\
&=& A^{(1)}_\gamma d\xi^{(1)\gamma} - V^{(1)}(\xi)dt,\end{eqnarray}
with $\gamma=1,2,\dots,(2N + 1)$ and
\begin{eqnarray}
\label{2040}
V^{(1)}&=&V^{(0)}|_{\Omega^{(0)}= 0},\nonumber\\
\xi^{(1)_\gamma} &=& (\xi^{(0)\theta},\eta),\\
A^{(1)}_\gamma &=&(A^{(0)}_\theta, \Omega^{(0)}).\nonumber
\end{eqnarray}
As a consequence, the first-iterative symplectic tensor is computed as
\begin{eqnarray}
\label{2050}
f^{(1)}_{\gamma\beta} = {\partial A^{(1)}_\beta\over \partial \xi^{(1)\gamma}}
-{\partial A^{(1)}_\gamma\over \partial \xi^{(1)\beta}}.
\end{eqnarray}
If this tensor is nonsingular, the iterative process stops and the Dirac's brackets
 among the phase space variables are obtained from the inverse matrix
 $(f^{(1)}_{\gamma\beta})^{-1}$ and, consequently, the Hamilton equation of
 motion can be computed and solved, as discussed in \cite{gotay}. It is well known
 that a physical system can be described at least classically in terms of a symplectic
 manifold $M$. From a physical point of view, $M$ is the phase space of the system while
 a nondegenerate closed 2-form $f$ can be identified as being the Poisson bracket. The
 dynamics of the system is  determined just specifying a real-valued function (Hamiltonian)
$H$ on the phase space, {\it i.e.}, one of these real-valued function
solves the Hamilton equation, namely,
\be \label{2050a1}
\iota(X)f=dH, \ee
and the classical dynamical trajectories of the
system in the phase space are obtained. It is important to mention
that if $f$ is nondegenerate, Eq. (\ref{2050a1}) has an unique
solution. The nondegeneracy of $f$ means that the linear map
$\flat:TM\rightarrow T^*M$ defined by $\flat(X):=\flat(X)f$ is an
isomorphism, due to this, the Eq.(\ref{2050a1}) is solved uniquely
for any Hamiltonian $(X=\flat^{-1}(dH))$. On the contrary, the
tensor has a zero-mode and a new constraint arises, indicating
that the iterative process goes on until the symplectic matrix
becomes nonsingular or singular. If this matrix is nonsingular,
the Dirac's brackets will be determined. In Ref. \cite{gotay}, the
authors consider in detail the case when $f$ is degenerate. The main idea of this embedding formalism is to introduce extra fields into the model in order to obstruct the solutions of the Hamiltonian equations of motion.
We introduce two arbitrary functions which are dependent on the original phase space and of WZ's variables, namely, $\Psi(a_i,p_i)$ and $G(a_i,p_i,\eta)$, into the first-order Lagrangian one-form as follows
\be
\label{2060a}
{\tilde{\cal L}}^{(0)}dt = A^{(0)}_\theta d\xi^{(0)\theta} + \Psi d\eta - {\tilde V}^{(0)}(\xi)dt,
\ee
with
\be
\label{2060b}
{\tilde V}^{(0)} = V^{(0)} + G(a_i,p_i,\eta),
\ee
where the arbitrary function $G(a_i,p_i,\eta)$ is expressed as an expansion in terms of the WZ field, given by
\begin{equation}
\label{2060}
G(a_i,p_i,\eta)=\sum_{n=1}^\infty{\cal G}^{(n)}(a_i,p_i,\eta),\,\,\,\,\,\,\,{\cal G}^{(n)}(a_i,p_i,\eta)\sim\eta^n\,,
\end{equation}
and satisfies the following boundary condition
\begin{eqnarray}
\label{2070}
G(a_i,p_i,\eta=0) = 0.
\end{eqnarray}
The symplectic variables were extended to also contain the WZ variable $\tilde\xi^{(0)\tilde\theta} = (\xi^{(0)\theta},\eta)$ (with ${\tilde\theta}=1,2,\dots,2N+1$) and the first-iterative symplectic potential becomes
\begin{equation}
\label{2075}
{\tilde V}^{(0)}(a_i,p_i,\eta) = V^{(0)}(a_i,p_i) + \sum_{n=1}^\infty{\cal G}^{(n)}(a_i,p_i,\eta).
\end{equation}
In this context, the new canonical momenta are
\be
{\tilde A}_{\tilde\theta}^{(0)} = \left\{\begin{array}{ll}
                                  A_{\theta}^{(0)}, & \mbox{with $\tilde\theta$ =1,2,\dots,2N}\\
                                  \Psi, & \mbox{with ${\tilde\theta}$= 2N + 1}
                                    \end{array}
                                  \right.
\ee
and the new symplectic tensor, given by
\begin{equation}
{\tilde f}_{\tilde\theta\tilde\beta}^{(0)} = \frac {\partial {\tilde A}_{\tilde\beta}^{(0)}}{\partial \tilde\xi^{(0)\tilde\theta}} - \frac {\partial {\tilde A}_{\tilde\theta}^{(0)}}{\partial \tilde\xi^{(0)\tilde\beta}},
\end{equation}
that is
\be
\label{2076b}
{\tilde f}_{\tilde\theta\tilde\beta}^{(0)} = \pmatrix{ { f}_{\theta\beta}^{(0)} & { f}_{\theta\eta}^{(0)}
\cr { f}_{\eta\beta}^{(0)} & 0}.
\ee

To sum up we have two steps: the first one is addressed to compute $\Psi(a_i,p_i)$ while the second one is dedicated to the calculation of $G(a_i,p_i,\eta)$. In order to begin with the first step, we impose that this new symplectic tensor (${\tilde f}^{(0)}$) has a zero-mode $\tilde\nu$, consequently, we get the following condition
\begin{equation}
\label{2076}
\tilde\nu^{(0)\tilde\theta}{\tilde f}^{(0)}_{\tilde\theta\tilde\beta} = 0.
\end{equation}
At this point, $f$ becomes degenerate and, in consequence, we introduce an obstruction to solve, in an unique way, the Hamilton equation of motion given in Eq.(\ref{2050a1}). Assuming that the zero-mode $\tilde\nu^{(0)\tilde\theta}$ is
\begin{equation}
\label{2076a}
\tilde\nu^{(0)}=\pmatrix{\mu^\theta & 1},
\end{equation}
and using the relation given in (\ref{2076}) together with (\ref{2076b}), we get a set of equations, namely,
\be
\label{2076c}
\mu^\theta{ f}_{\theta\beta}^{(0)} + { f}_{\eta\beta}^{(0)} = 0,
\ee
where
\be
{ f}_{\eta\beta}^{(0)} =  \frac {\partial A_\beta^{(0)}}{\partial \eta} - \frac {\partial \Psi}{\partial \xi^{(0)\beta}}.
\ee
The matrix elements $\mu^\theta$ are chosen in order to disclose a desired gauge symmetry. Note that in this formalism the zero-mode $\tilde\nu^{(0)\tilde\theta}$ is the gauge symmetry generator. At this point, it is worth to mention that this characteristic is important because it opens up the possibility to disclose the desired hidden gauge symmetry from the noninvariant model. It awards to the symplectic embedding formalism some power to deal with noninvariant systems. From relation (\ref{2076}) some differential equations involving $\Psi(a_i,p_i)$ are obtained, Eq. (\ref{2076c}), and after a straightforward computation, $\Psi(a_i,p_i)$ can be determined.

In order to compute $G(a_i,p_i,\eta)$ in the second step, we impose that no more constraints arise from the contraction of the zero-mode $(\tilde\nu^{(0)\tilde\theta})$ with the gradient of the potential ${\tilde V}^{(0)}(a_i,p_i,\eta)$. This condition generates a general differential equation, which reads as
\begin{widetext}
\begin{eqnarray}
\label{2080}
\tilde\nu^{(0)\tilde\theta}\frac{\partial {\tilde V}^{(0)}(a_i,p_i,\eta)}{\partial{\tilde\xi}^{(0)\tilde\theta}}\,&=&\, 0,\\
\mu^\theta \frac{\partial {V}^{(0)}(a_i,p_i)}{\partial{\xi}^{(0)\theta}} + \mu^\theta \frac{\partial {\cal G}^{(1)}(a_i,p_i,\eta)}{\partial{\xi}^{(0)\theta}} 
\,+\, \mu^\theta\frac{\partial {\cal G}^{(2)}(a_i,p_i,\eta)}{\partial{\xi}^{(0)\theta}} + &\dots& 
\,+\,\frac{\partial {\cal G}^{(1)}(a_i,p_i,\eta)}{\partial\eta} + \frac{\partial {\cal G}^{(2)}(a_i,p_i,\eta)}{\partial\eta} + \dots = 0\;\;, \nonumber
\end{eqnarray}
\end{widetext}
that allows us to compute all correction terms ${\cal G}^{(n)}(a_i,p_i,\eta)$ in order of $\eta$. Note that this polynomial expansion in terms of $\eta$ is equal to zero, consequently, whole coefficients for each order in $\eta$ must be null identically. In view of this, each correction term in order of $\eta$ is determined. For a linear correction term, we have
\begin{equation}
\label{2090}
\mu^\theta\frac{\partial V^{(0)}(a_i,p_i)}{\partial\xi^{(0)\theta}} + \frac{\partial{\cal
 G}^{(1)}(a_i,p_i,\eta)}{\partial\eta} = 0.
\end{equation}
For a quadratic correction term, we get
\begin{equation}
\label{2095}
{\mu}^{\theta}\frac{\partial{\cal G}^{(1)}(a_i,p_i,\eta)}{\partial{\xi}^{(0)\theta}} + \frac{\partial{\cal G}^{(2)}(a_i,p_i,\eta)}{\partial\eta} = 0.
\end{equation}
From these equations, a recursive equation for $n\geq 2$ is proposed as
\begin{equation}
\label{2100}
{\mu}^{\theta}\frac{\partial {\cal G}^{(n - 1)}(a_i,p_i,\eta)}{\partial{\xi}^{(0)\theta}} + \frac{\partial{\cal
 G}^{(n)}(a_i,p_i,\eta)}{\partial\eta} = 0,
\end{equation}
that allows us to compute the remaining correction terms in order of $\eta$. This iterative process is successively repeated until (\ref{2080}) becomes identically null, consequently, the extra term $G(a_i,p_i,\eta)$ is obtained explicitly. Then, the gauge invariant Hamiltonian, identified as being the symplectic potential, is obtained as
\begin{equation}
\label{2110}
{\tilde{\cal  H}}(a_i,p_i,\eta) = V^{(0)}(a_i,p_i) + G(a_i,p_i,\eta),
\end{equation}
and the zero-mode ${\tilde\nu}^{(0)\tilde\theta}$ is identified as being the generator of an infinitesimal gauge transformation, given by
\begin{equation}
\label{2120}
\delta{\tilde\xi}^{\tilde\theta} = \varepsilon{\tilde\nu}^{(0)\tilde\theta},
\end{equation}
where $\varepsilon$ is an infinitesimal parameter.

\section{The massive Carrol-Field-Jackiw model}

The study of both Lorentz and gauge invariance in variations of
Maxwell's model is of strong theoretical
\cite{Colladay,Coleman,Jackiw,Andrianov,Adam,Perez,Baeta} and
experimental \cite{Carroll} interest and great relevance in
practical applications as the quantum Hall effect \cite{Girvin}
and high-$T_c$ superconductivity \cite{Polyakov}.

The construction of dual equivalent and a gauge-invariant version of the Maxwell modified
theory will now be accomplished in the symplectic framework.
The DEM introduces extra variables which enlarge the
phase space \cite{aon} furnishing a dual equivalent action of the first one, and, furthermore restore the
gauge-invariance of the theory.

\subsection{The symplectic analysis}

In this section, the MCS theory in four dimensions will be
analyzed from the symplectic point of view \cite{amnot}. Let us consider the
massive Maxwell-Chern-Simons Lagrangian in four dimensions
\cite{Carroll,helayel}
\be \label{01} {\cal L}=-\frac{\beta}{4}F_{\mu\nu}F^{\mu\nu}
+\frac{m^2}{2}A_\mu A^\mu -\frac{1}{4}p_\alpha A_\beta
\epsilon^{\alpha\beta\mu\nu}F_{\mu \nu}, \ee
where $p$ is an external constant four-vector, which selects a preferred direction in space-time for each Lorentz frame. 
This term couples the electromagnetic field to an four-vector $p_{\alpha}$ \cite{Carroll}.
Now, following the symplectic method the zeroth-iterative first-order Lagrangian
one-form is written as
\begin{widetext}
\ba \label{02} 
{\cal L}\,=\,\pi^i \dot A_i &+&A_0\left( \partial^i
\pi_i +m^2 A_0 +\frac{1}{4}p^i\epsilon_{ijk} F^{ik}\right)
\,+\,\frac{1}{2\beta}p_i A_j \pi_k \epsilon^{ijk} 
+\frac{1}{2\beta}\pi_i \pi^i
\,-\,\frac{\beta}{4}F_{ij}F^{ij} -\frac{1}{2}m^2 A_0 A^0 
\,+\,\frac{1}{2}m^2A_iA^i \nonumber \\
&+&\frac{1}{8\beta}p^iA_j \(p_jA^i -p_i A^j \) 
\,-\, \frac{1}{4}p^0 A^i \epsilon_{ijk}F^{jk} \;\;,
\ea
\end{widetext}
with the canonical momentum $\pi_i$  given by
\ba \label{03} \pi_i &=& -\beta F_{0i} -\frac{1}{2}p^jA^k
\epsilon_{ijk}\nonumber\\
&=& -\beta\left(\dot A_i -\partial_i A_0\right) -\frac{1}{2}p^jA^k
\epsilon_{ijk}. \ea

The symplectic fields are $\xi^{(0)\alpha}\,=\,\(A^i ,\pi^i ,A^0 \)$
and the zeroth-iterative symplectic matrix is
\be \label{04} f^{(0)}=
\left(%
\begin{array}{ccc}
  0 & -\delta^i_j & 0 \\
  \delta^j_i & 0 & 0 \\
  0 & 0 & 0 \\
\end{array}%
\right)\delta(x-y) \ee
which is a singular matrix. It has a zero-mode that generates the
following constraint
\be \label{05} \Omega(x) = \partial_i \pi^i(x) + m^2 A_0 (x)
+\frac{1}{4}p^iF^{jk}\epsilon_{ijk}, \ee
identified as being the Gauss law. Bringing back this constraint
into the canonical part of the first-order Lagrangian density
${\cal L}^{(0)}$ using a Lagrangian multiplier $\(\zeta\)$, the
first-iterated Lagrangian density, written in terms of the following
symplectic fields $\xi^{(1)\alpha} =\(A^i,\pi^i,A^0,\zeta\)$ is
obtained as
\begin{widetext}
\ba \label{06} 
{\cal L}^{(1)} &=&\pi^i \dot A_i +\Omega\dot \zeta
\,+\,\frac{1}{2\beta}p_i A_j \pi_k \epsilon^{ijk}
\,+\,\frac{1}{2\beta}\pi_i \pi^i 
\,-\,\frac{\beta}{4}F_{ij}F^{ij}
-\frac{1}{2}m^2 A_0 A^0 \,+\,\frac{1}{2}m^2A_iA^i \nonumber\\
&+&\frac{1}{8\beta}p^iA_j \(p_jA^i -p_i A^j \) 
\,-\, \frac{1}{4}p^0 A^i \epsilon_{ijk}F^{jk}. 
\ea
\end{widetext}
The first-iterated symplectic matrix is obtained as being
\ba 
& &\label{07} f^{(1)}= \nonumber\\
& & \!\!\!\!\!\!\!\left(
\begin{array}{cccc}
  0 & -\delta_j^i \delta(x-y) & 0 & f_{A^i \zeta} \\
  \delta_i^j\delta(x-y) & 0 & 0 & \partial_i^y\delta(x-y) \\
  0 & 0 & 0 & m^2\delta(x-y) \\
  f_{\zeta A^j} & -\partial_j^y\delta(x-y) & -m^2\delta(x-y) & 0 \\
\end{array}\right) \nonumber\\
\mbox{}
\ea
where
\be \label{08} f_{A^i \zeta}=-\frac{1}{2}p^n\partial_y^m
\delta(x-y) \epsilon_{nim}. \ee
This matrix is nonsingular and, as settle by the symplectic
formalism, the Dirac brackets among the phase space fields are
acquired from the inverse of the symplectic matrix, namely,
\ba \label{09} \{A^i(x),A^j(y)\}^*&=&0,\nonumber\\
\{A^i(x),\pi^j(y)\}^* &=& \delta^{ij}\delta(x-y),\nonumber\\
\{A^0(x),A^j(y)\}^* &=& \frac{1}{m^2}\partial_x^j \delta(x-y),\\
\{A^0(x),\pi^j(y)\}^* &=&
\frac{1}{2m^2}\epsilon^{lij}p_l\partial_i^x\delta(x-y).\nonumber
\ea

As we said above, the basic symplectic analysis was the first step of the DEM.  The next step is to introduce the WZ fields in order to proceed with the dualization.  This will be done in the next section.

\subsection{The Dual Equivalent Model}

Now the phase space will be extended with the introduction of the WZ
fields. In order to start, we change the Lagrangian,
Eq. (\ref{02}), introducing two arbitrary functions
$\psi\equiv\psi\(A^i,\pi^i,A^0, \eta\)$ and $G\equiv
G\(A^i,\pi^i,A^0, \eta\)$ with the WZ field, namely,
\be \label{11} \tilde{\cal L}^{(0)} =\pi_i \dot A^i + \psi\dot\eta
-\tilde{V}^{(0)}, \ee
where the symplectic potential is
\begin{widetext}
\ba \label{12} 
\tilde {V}^{(0)} &=&-A_0\left( \partial^i \pi_i
+m^2 A_0 \,+\,\frac{1}{4}p^i\epsilon_{ijk} F^{ik}\right) 
\,-\,\frac{1}{2\beta}p_i A_j \pi_k \epsilon^{ijk}
-\frac{1}{2\beta}\pi_i \pi^i +\frac{\beta}{4}F_{ij}F^{ij}
\,+\,\frac{1}{2}m^2 A_0 A^0 -\frac{1}{2}m^2A_iA^i \nonumber \\
&-&\,\frac{1}{8\beta}p^iA_j \(p_jA^i -p_i A^j \) 
\,+\, \frac{1}{4}p^0 A^i \epsilon_{ijk}F^{jk} \,+\, G, 
\ea
\end{widetext}
and $G$ is a function expressed as
\be \label{13} G(A^i,\pi^i, A^0, \eta)=\sum_{n=1}^\infty {\cal
G}^{n} \,\,\, \mbox{with} \,\,\, {\cal G}^{n} \propto\eta^n .\ee
The arbitrary function satisfies the following boundary condition,
\be \label{14} G\(A^i,\pi^i,A^0,\eta=0 \) =0. \ee

The extended symplectic field are $\tilde \xi^{(0)}
=\(A^i,\pi^i,A^0,\eta\)$ and the corresponding matrix is
\be\label{15} \tilde{f}^{(0)}=\left(\begin{array}{cccc}
  0 & -\delta_j^i\delta(x-y) & 0 & \frac{\delta\psi(y)}{\delta A^i(x)} \\
  \delta_i^j\delta(x-y) & 0 & 0 & \frac{\delta\psi(y)}{\delta \pi^i(x)} \\
  0 & 0 & 0 & \frac{\delta\psi(y)}{\delta A^0(x)} \\
  -\frac{\delta\psi(x)}{\delta A^j(y)} & -\frac{\delta\psi(x)}{\delta \pi^j(y)} & -\frac{\delta\psi(x)}{\delta A^0(y)} & 0 \\
\end{array}
\right). \ee
This singular matrix has a zero-mode, which can be settle conveniently as
\be \label{16} \tilde \nu = \pmatrix{\partial^i & 0 & \partial^0 &
1 }. \ee
Contracting this zero-mode with the symplectic matrix above, a set
of differential equation can be obtained as
\ba \label{17} 
& &\int dx\;\; \( \frac{\delta\psi(y)}{\delta
A^i(x)}\)\,=\,0, \nonumber\\
& &\int dx\;\; \( \delta_i^j\partial^x_j\delta(x-y) +
\frac{\delta\psi(y)}{\delta\pi^i(x)}\)\,=\,0,\nonumber\\
& &\int dx\;\; \( \frac{\delta\psi(y)}{\delta A^0(x)}\)\,=\,0, \\
& &\int dx\;\; \(-\partial_x^j\frac{\delta\psi(x)}{\delta A^j(y)}
-\partial_x^0\frac{\delta\psi(x)}{\delta A^0(y)}\)\,=\,0. \nonumber
\ea
After a straightforward computation, we get
\be \label{18} \psi(x) = -\partial^i \pi_i(x). \ee
Then, the Lagrangian becomes
\be \label{19} \tilde{\cal L}^{(0)} =\pi_i \dot A^i
-\partial^i\psi_i\dot\eta -\tilde{V}^{(0)}\,. \ee

After this point, we begin with the final step of the symplectic
embedding formalism. To this end, we impose that the contraction
of the zero-mode, Eq. (\ref{16}), with the gradient of the
symplectic potential generates an identically null result, namely,
\be\label{20} \int dy\;\; \tilde \nu^{(0)}(x)\frac{\delta\tilde
V^{(0)}(y)}{\delta\tilde\xi^{(0)}(x)}=0\,\,. \ee
From this condition, the following general differential equation
is obtained,
\ba \label{21} 
& & \int dy \,\[\partial_x^i \(\frac{\delta\tilde{V}^{(0)}(y)}{\delta A^i(x)}\)+
\partial_x^0 \(\frac{\delta\tilde{V}^{(0)}(y)}{\delta A^0(x)}\) \right. \nonumber \\
& & \left.+\; 1.\(\sum_{n=1}^{\infty}\frac{\delta{\cal G}^{(n)}(y)}{\delta \eta(x)}\)
\]=0,
\ea
where the relation given in (\ref{13}) was used. This allows
the computation of the whole correction terms in order of $\eta$.
For linear correction term $\({\cal G}{(1)}(x)\)$, we get
\begin{widetext}
\ba \label{22} 
{\cal G}^{(1)}\,&=&\,\[{1\over2}p^i\partial^jA^0(x)\epsilon_{ijl}
\,+\,{1\over2}p^iF^{0k}(x)\epsilon_{ilk}
-\beta\partial^iF_{il}(x)\,-\,m^2A_l(x)+
{1\over4}p^0F^{jk}(x)\epsilon_{jkl}-{1\over2}p^0\partial^j
A^i\epsilon_{ijl}\]\partial^l \eta \nonumber\\
&-&\[\partial^i\pi_i(x) +m^2A_0(x)
+{1\over4}p^iF^{jk}(x)\epsilon_{ijk}\]\partial^0\eta.  
\ea
\end{widetext}

For the quadratic correction term, we have
\ba \label{23} 
& &\int dy \[
\partial_x^i \(\frac{\delta{\cal G}^{(1)}(y)}{\delta A^i(x)}\)+
\partial_x^0 \(\frac{\delta{\cal G}^{(1)}(y)}{\delta A^0(x)}\) \right. \nonumber \\
& & \left.+\, 1.\(\frac{\delta{\cal G}^{(2)}(y)}{\delta \eta(x)}\)
\]=0,
\ea
with the following solution,
\be\label{24}{\cal
G}^{(2)}=-{m^2\over2}\partial_i\eta\partial^i\eta-
{m^2\over2}\partial_0\eta\partial^0\eta. \ee

Note that the second-order correction term has dependence only on the
WZ field, thus all the correction terms ${\cal G}^{(n)}$ for
$n\geq 3$ are null. Then, the gauge invariant first-order
Lagrangian density is given by
\begin{widetext}
\ba \label{25} 
\tilde{\cal L} &=& {\cal L} \,+\,\[ m^2 A_k
+\beta\partial^0 F_{0k} +\beta\partial^i F_{ik} -p^0\partial^iA^j
\epsilon_{ijk} 
\,+\, p^iF^{j0}\epsilon_{ijk}\]\partial^k\eta\,+\,\[m^2 A_0 +\beta\partial^i F_{io}-
p^i\partial^jA^k\epsilon_{ijk}\]\partial^0\eta \nonumber\\
&+&{m^2\over2}\(\partial_i\eta\partial^i\eta
\,+\,\partial_0\eta\partial^0\eta\), 
\ea
\end{widetext}
where $\cal L$ is given in (\ref{01}). We may recognize the
Noether current in Eq.(\ref{25}) as
\ba\label{26}
J_k &=&m^2 A_k +\beta\partial^0 F_{0k}
+\beta\partial^i F_{ik} -p^0\partial^iA^j \epsilon_{ijk}
+p^iF^{j0}\epsilon_{ijk}, \nonumber\\
J_0 &=&m^2 A_0 +\beta\partial^i F_{io}-
p^i\partial^jA^k\epsilon_{ijk}. \ea
So, we can write $\tilde {\cal L}$ as
\be\label{27} \tilde{\cal L} = {\cal L} +J_{\mu}\partial^{\mu}\eta
+ {m^2\over2}\partial_{\mu}\eta\partial^{\mu}\eta . \ee
Solving for $\partial_\mu \eta$ we get that
\be\label{28} J_\mu +m^2\partial_{\mu}\eta=0. \ee
Plugging this back into (\ref{27}), we obtain the remarkable
gauge-invariant theory
\begin{widetext}
\ba\label{29} 
\tilde{\cal L} &=&{\beta\over4}F_{\mu\nu}F^{\mu\nu}+
{1\over2}\epsilon_{\alpha\beta\nu\mu}p^{\alpha}\(\partial^{\beta}A^{\nu}\)A^{\mu} 
\,-\,{1\over{2m^2}}\[\epsilon_{\alpha\beta\nu\mu}p^{\alpha}\(\partial^{\beta}A^{\nu}\)\]^2 
\,-\,{\beta^2\over{2m^2}} 
\[\partial_\mu F^{\mu\nu}\]^2 \nonumber\\
&+&{\beta\over{m^2}}\;\epsilon_{\alpha\beta\nu\mu}p^{\alpha}\(\partial^{\beta}A^{\alpha}\)\(\partial_{\rho}F^{\rho\mu}\).
\ea
\end{widetext}
which is the same result obtained in \cite{helayel}, using the NDM, i.e., the action (\ref{29}) is the dual equivalent to the action (\ref{01}).  We see that as the zero-mode given in (\ref{16}) is arbitrary, any other zero-mode for the symplectic matrix (\ref{15}) will bring us a new dual equivalent action.  We believe that this is one great advantage of the DEM when we confront this with the NDM.  Since our objective in this work is to prove that the DEM can produce dual equivalent actions as the NDM, we used the zero-mode which reproduce precisely the action obtained in \cite{helayel}, namely, the Eq. (\ref{29}).

To complete the comparison between both methods, as well known from the symplectic formalism literature, the zero-mode is the generator of the infinitesimal gauge transformations, since Eq. (\ref{29}) is also the gauge invariant version of (\ref{01}).  We believe that this constitute another good point in favor for the DEM.  So, using the zero mode,
Eq. (\ref{16}), as the generator of the infinitesimal gauge
transformations given by  $\delta{\cal O}=\epsilon \tilde \nu^{(0)}$, we have
\ba\label{30} \delta A_i &=& -\partial_i \epsilon,\nonumber\\
\delta \pi_i &=&0,\nonumber\\
\delta A_0 &=& -\partial_0 \epsilon, \\
\delta \eta &=& \epsilon, \nonumber \ea
where $\epsilon$ is an infinitesimal time-dependent parameter.

It is important to notice that more than one WZ symmetry will be unveiled (see \cite{fluidos} for a review), showing that the studied model does not have a unique WZ gauge-invariant description \cite{henrique}, but a family of dynamically equivalent WZ gauge-invariant representations.  This can allow an interesting discussion concerning both obvious symmetry (phase symmetry) and hidden symmetry (Galileo antiboost invariance) of the studied model.   For example, the additional symmetries found in \cite{bazeia} were investigated in \cite{fluidos} from the symplectic embedding point of view.  Indeed, the global status of these symmetries will be lifted to local.  We believe that this property of unveiling a whole family of symmetries and consequently a whole family of equivalent actions is the great advantage of the DEM in comparison with NDM.

\section{The self-dual model minimally coupled to $U(1)$ charged bosonic matter}

Let us consider next the case of the self-dual model minimally coupled to U(1) charged bosonic matter which is
described by the following Lagrangian density \cite{ainrw,GMdS},

\begin{equation}
\label{PB120}
{\cal{L}}_{min}^{(0)}={\cal L}_{SD} + {\cal L}_{int}+{\cal{L}}_{KG},
\end{equation}
where

\begin{eqnarray}
{\cal L}_{SD} &=& \frac{m^{2}}{2}f^{\mu}f_{\mu}
-\frac{m}{2}\varepsilon^{\mu\nu\rho}f_{\mu}\partial_{\nu}f_{\rho}\nonumber\\
{\cal L}_{int} &=& -ef_{\mu}J^{\mu}+e^{2}f^{\mu}f_{\mu}\phi^{*}\phi\nonumber\\
{\cal{L}}_{KG}&=&\partial_{\mu}\phi^{*}\partial^{\mu}\phi-M^{2}\phi^{*}\phi
\, ,
\end{eqnarray}
are the self-dual, interaction and Klein-Gordon Lagrangian for a vector
field and a massive U(1) charged, complex scalar field, respectively.  Here,
\begin{equation}
\label{PB130}
J^{\mu}=i(\phi^{\ast}\partial^{\mu}\phi-\partial^{\mu}\phi^{\ast}\phi),
\end{equation}
is the global Noether current associated to a U(1) phase transformation.

The canonical momenta are
\ba
\pi_j &=& {m\over2}\,\epsilon_{ij}\,f^i \nonumber \\
p&=&-i\,e\,f_0\phi^*\,+\,\pa_0\,\phi^* \\
P&=&i\,e\,f_0\,\phi\,+\,\pa_0\,\phi\,\,.
\ea

Hence, we have the following Lagrangian,
\be
{\cal L}\,=\,\pi^i\,f_i\,+\,p\,\dot{\phi}\,+\,P\,\dot{\phi}^*\,-\,V^{(0)}
\ee
where
\begin{widetext}
\ba
V^{(0)}\,&=&\,{1\over2}\,m\epsilon^{ijk}\,F_i\pa_j\,f_k
\,+\,{m\over2}\epsilon^{0ij}\,f_0\,\pa_i\,f_j\,+\,{m\over2}\epsilon^{ij0}\,f_i\,\pa_j\,f_0\,+\,p\,P\,+i\,e\,f_0\,(P\phi^*\,-\,p\,\phi)\,+\,i\,e\,f_i\,(\phi^*\pa^i\phi\,-\,\phi\,\pa^i\phi^*) \nonumber \\
&\,+&\,M^2\,\phi\,\phi^*\,-\,{1\over2}\,m^2\,f^{\mu}\,f_{\mu}\,-\,\pa_i\,\phi^*\,\pa^i\,\phi\,-\,e^2\,f^2_i\,\phi^*\phi\,\,.
\ea
\end{widetext}

Considering a convenient zero-mode like,
\be \l{modozero2}
\tilde{\nu}^{(0)}\,=\,(\pa^l_x\,\,\,0\,\,\,\pa^0_x\,\,\,0\,\,\,0\,\,\,0\,\,\,0\,\,\,0\,\,\,1)
\ee
we have that
\be
\p\,=\,-\,\pa^i\,\pi_i \,\,.
\ee

Contracting the zero-mode (\ref{modozero2}) with the symplectic potential we have that,
\begin{widetext}
\ba
& &\int dy\,\left[\,{m\over2}\,\epsilon_{ijk}\,\pa^i_x\,\dirac\,\pa^j\,f^k
\,+\,{m\over2}\,\epsilon_{ijk}\,f^i\,\pa^j_y\,\pa^{k}\dirac
\,+\,{m\over2}\,\epsilon_{0ij}\,f^0\,\pa^i\,\pa^j_x\,\dirac
\,+\,{m\over2}\,\epsilon_{ij0}\,\pa^j\,f^0\,\pa^i_x\,\dirac \, \right. \nonumber \\
& & \left. +\,i\,e\,\pa_x^i\dirac\,(\phi^*\,\pa_i\phi\,-\,\phi\,\pa_i\,\phi^*)\,-\,m^2\,f_i\,\pa_x^i\,\dirac\,-\,e^2\,f_i\,\pa^i_x\,\dirac\,\phi^*\,\phi\,+\,{m\over2}\,\epsilon_{0ij}\pa^0_x\,\dirac\pa^i\,f^i\, \right. \\
& & \left. \,+\,
{m\over2}\,\epsilon_{ij0}\,f^i\,\pa^j\pa^0_x\,\dirac\,+\,i\,e\,\pa^0_x\,\dirac\,(P\phi^*\,-\,p\,\phi)\,-\,
m^2\,f_0\,\pa^0_x\,\dirac\,+\,\frac{\delta\,{\cal G}^{(0)}(y)}{\delta \eta(x)}\right] = 0 \,\,,\nonumber
\ea
\end{widetext}
with the following solution,
\ba
{\cal G}^{(0)}&=&\,(e\,J_i\,-\,m^2\,f_i\,-\,2\,e^2\,f_i\,\phi\,\phi^*)\,\pa^i\,\eta \nonumber \\
&+&(e\,J_0\,-\,m^2\,f_0\,-\,2\,e^2\,f_0\,\phi\phi^*)\,\pa^0\eta \\
&+&m\epsilon_{ijk}\,\pa^i\,\eta\,\pa^jf^k\,+\,m\epsilon_{ij0}\,\pa^i\eta\,\pa^j\,f^0\, \nonumber \\
&+&\,m\epsilon_{0ij}\,\pa^0\,\eta\,\pa^i\,f^j \,\,.\nonumber
\ea

With this result we have a new symplectic potential,
\be
\tilde{V}^{(1)}\,=\,\tilde{V}^{(0)}\,+\,{\cal G}^{(0)}\,+\,\sum^{\infty}_{n=1}\,{\cal G}^{(n)}
\ee
and again, a contraction of this result with the zero-mode gives,
\ba
& &\int dy\,\left[ \,-m^2\,\pa_x^i\,\dirac\,\pa_i\,\eta\,-\,2\,e^2\,\pa_x^i\,\dirac\,\phi\phi^*\,\pa_i\,\eta \right.\nonumber \\
& & \left.\,-\,m^2\,\pa_x^0\,\dirac\,\pa_0\,\eta\,-\,2\,e^2\,\pa^0_x\,\dirac\,\phi\phi^*\,\pa^0\,\eta \right.\nonumber \\
& & \left. \,+\,\frac{\delta\,{\cal G}^{(0)}(y)}{\delta \eta(x)}\right]\,=\,0\,\,, 
\ea
namely,
\be
{\cal G}^{(1)}\,=\,-\,{m^2 \over2}\,\pa_{\mu}\eta\pa^{\mu}\eta\,-\,e^2\phi\,\phi^*\,\pa_{\mu}\,\pa^{\mu}\eta\,\,,
\ee
and substituting this result in the symplectic potential we have a new one which is,
\ba
& &\tilde{V}\,=\,V^{(0)}\,+\,(e\,J_{\mu}\,-\,m^2\,f_{\mu}\,-\,2\,e^2\,f_{\mu}\,\phi\,\phi^*\,)\,\pa^{\mu}\eta \nonumber \\
&+&m\epsilon_{ijk}\,\pa^i\,\eta\,\pa^j\,\f^k\,+\,m\,\epsilon_{ij0}\,\pa^i\,\eta\,\pa^j\,f^0\,+\,m\epsilon_{0ij}\,\pa^0\,\eta\,\pa^i\,f^j \nonumber \\
&-&{m^2\over2}\pa_{\mu}\eta\,\pa^{\mu}\eta\,-\,e^2\,\phi\,\phi^*\,\pa_{\mu}\eta\,\pa^{\mu}\eta\,\,.
\ea

After a little algebra we can rewrite the Lagrangian as,
\be
\tilde{{\cal L}}\,=\,{\cal L}\,-\,K_{\mu}\,\pa^{\mu}\eta\,+\,{1\over2}(m^2\,+\,2e^2\phi\phi^*\,)\pa_{\mu}\eta\pa^{\mu}\eta
\ee
where
\be
K_{\mu}\,=\,e\,J_{\mu}\,-\,m^2\,f_{\mu}\,-\,2e^2\,f_{\mu}\,\phi\phi^*\,+\,m\epsilon_{\mu\nu\rho}\,\pa^{\nu}f^{\rho}
\ee
and conveniently, let us define $\pa_{\mu}\eta$ as an external field $B_{\mu}$, and we can write
\be
\tilde{{\cal L}}\,=\,{\cal L}\,-\,K_{\mu}\,B^{\mu}\,+\,{1\over2}(m^2\,+\,2e^2\phi\phi^*)\,B_{\mu}B^{\mu}\,\,.
\ee

Solving the equations of motion for $K_{\mu}$ we have,
\be
\tilde{{\cal L}}\,=\,{\cal L}\,-\,{1\over2}(m^2\,+\,2e^2\phi\phi^*)\,B_{\mu}B^{\mu}\,\,,
\ee
and eliminating the WZ fields we can work algebraically to obtain the final equivalent dual theory as,
\ba
\tilde{{\cal L}}\,&=&\,{\cal L}_{KG}\,-\,\frac{m^2}{4{\mu}^2}\,F_{\mu\nu}^2\,+\,{m\over2}\epsilon_{\mu\nu\rho}\,A^{\mu}\,\pa^{\nu}A^{\rho} \nonumber \\
&\,-\,&\frac{e^2}{2{\mu}^2}\,J^2\,-\,\frac{m^2}{{\mu}^2}\,F_{\mu}J^{\mu}\,\,,
\ea
which is the same result found in \cite{ainrw}.   We can aver that this result has the same construction as in the fermionic case, i.e., where a fermionic matter field is coupled to the self-dual field.
The minimal coupling is replaced by a non-minimal magnetic coupling and the presence of the Thirring-like current-current term \cite{ainrw}.  Notice that now the coefficient of the Maxwell term is field dependent.
Abelian Higgs model carries this kind of structure, however, with an anomalous magnetic interaction \cite{ln}.

\section{Conclusions}

Following the idea of Faddeev and Shatashvilli, the dual embedding formalism is based on a contemporary framework that handles constrained models which is called symplectic formalism.  In the introduction we described the favorable points in favor of it.  The effectiveness of the method was demonstrated through several papers in the literature.  We can say that the investigation of how to obtain dual equivalent actions to systems with many degrees of freedom is quite desirable since these systems live in a world permeated with non-perturbative features that need special and difficult treatment.

For a review \cite{amnot} we promote the dualization of the gauge-invariant Maxwell theory modified by the
introduction of an explicit massive (Proca) term and a topological but not Lorentz-invariant term \cite{Carroll,helayel}.   Afterwards, this noninvariant theory was reformulated as a gauge-invariant/dual equivalent theory via DEM where the gauge-invariance broken was restored. 

In this work we used the dual embedding method to dualize the self-dual model minimally coupled to $U(1)$ charged bosonic matter.  The final result shows the same structure as the fermionic case, i.e., where a fermionic matter field is coupled to the self-dual field \cite{ainrw}.  The minimal coupling is replaced by a non-minimal magnetic coupling and the presence of the Thirring-like current-current term.  The coefficient of the Maxwell term now is field dependent.   It can be shown that this kind of construction is important in Abelian Higgs model with anomalous magnetic interaction \cite{ln}.



\section{ Acknowledgments}

EMCA would like to thank the hospitality and kindness of the Dept. of Physics of the Federal University of Juiz de Fora where part of this work was done.
The authors would like  to thank CNPq, FAPEMIG and FAPERJ (Brazilian financial agencies) for financial support.

\end{document}